\documentstyle[preprint,aps,prb]{revtex}

\begin{document}
\draft

\date{\today}
\title{Spin effects in a confined 2DEG: Enhancement of the $g$-factor,
       spin-inversion states and their far-infrared absorption.}
\author{Vidar Gudmundsson$^1$ and Juan Jos\'e Palacios$^2$}
\address{$^1$Science Institute, University of Iceland, Dunhaga 3,
         IS-107 Reykjavik, Iceland.\\
    $^2$Department of Physics, Swain Hall West 117, Indiana University
         at Bloomington, IN 47405, USA.}
%
%\tighten

\preprint{IUCM 95-010}
\maketitle
\abstract{We investigate several spin-related phenomena
          in a confined two-dimensional electron gas (2DEG)
          using the Hartree-Fock approximation for the
          mutual Coulomb interaction of the electrons.
          The exchange term of the interaction causes a large
          splitting of the spin levels whenever the
          chemical potential lies within a Landau band (LB).
          This splitting can be reinterpreted
          as an enhancement of an effective $g$-factor, $g^*$.
          The increase of $g^*$ when a LB is half filled
          can be accompanied by a spontaneous formation of
          a static spin-inversion state (SIS) whose details depend on
          the system size. The coupling of the states of higher
          LB's into the lowest band by the Coulomb interaction
          of the 2DEG is essential for the SIS to occur.
          The far-infrared absorption
          of the system, relatively insensitive to the spin splitting,
          develops clear signs of the SIS.
}

\section{Introduction}
The effects of the exchange interaction on the appearance
of macroscopic spin structures have been
studied in semiconductor microstructures in reduced dimensions
by several researchers both theoretically and in experiments.
The enhancement of the effective $g$-factor, $g^*$, of a
two-dimensional electron gas (2DEG) in the quantum Hall regime has been
reviewed by Ando, Fowler, and Stern.\cite{Ando82:437}
For the unbounded 2DEG
Ando and Uemura \cite{Ando74:1044} presented a model
where the broadening of the Landau levels due to
impurity scattering is treated in the self-consistent
Born approximation (SCBA). The dielectric function is calculated
with the inclusion of the lowest order exchange energy of the
screened Coulomb interaction in the self-energy of the electrons.
For a strong magnetic field the overlapping of Landau levels with
different indices is neglected.

The enhancement of $g^*$
can lead to a spin polarization of the 2DEG at certain values of the filling
factor $\nu$, and, in addition, the exchange interaction
can lead to the spontaneous formation of spin-density \cite{Gruener:xx} or
charge-density waves \cite{Gerhardts81:1339,Yoshioka:4986}.
The onset of a spin-density wave state in a parabolic quantum well
has been studied by Brey and Halperin using a modified Hartree-Fock
approximation (HFA)
with a point-contact exchange interaction. They find a divergence
of the electric susceptibility in the presence of a
magnetic field of intermediate strength parallel to the quantum well
and an infinitesimal fictitous magnetic field perpendicular to
the quantum well.\cite{Brey89:11634} This spin-density wave state
has a wavevector along the quantum well parallel to the intermediate
magnetic field and occurs only when the quantum
well is wide enough and the exchange interaction has a strength
larger than a critical value. The calculated optical properties of a
$\delta$-doped quantum well in the HFA
due to spin- and charge-density excitations have been found to be
in good agreement with experiments,\cite{Luo93:11086} as well as those
of donor states in 2DEG in strong magnetic fields.\cite{Hawrylak94:2943}
In the quantum well Hembree {\em et al.}\cite{Hembree93:9162}
discovered abrupt spin polarization of the system
at high magnetic fields and a spin-inversion regime where the net spin
alignment strongly varies across the well. They studied the effect
in different approximation schemes and in the presence of
impurity scattering. Recently the effects of the $g$-factor enhancement
on various transport coefficients has been reported by the
same group.\cite{Hembree94:15197}

As to microstructures of further reduced dimensionality
the spontaneous polarization of
of an array of quantum dots into a ferroelectric or antiferroelectric state
has been investigated by Kempa, Broido, and Bakshi.\cite{Kempa:9343}
The spin degree of freedom together with the exchange interaction
and correlation effects have also been found to be essential
to model few electrons in a single quantum dot in magnetic
field.\cite{Pfannkuche93:2244,Pfannkuche93:6,Hawrylak93:3347,Palacios94:5760,Ferconi94:14722}

In this paper we are concerned with the spin-related phenomena associated
with the exchange interaction that can occur in quantum dots with a large
number of electrons.  We study the spin splitting of Landau bands
(LB's) due to the enhancement of $g^*$,
and the formation of a spin-inversion state (SIS)
in a strictly two-dimensional {\em finite size} electron system in a
perpendicular magnetic field of intermediate strength.
The system size is chosen to be of the order of
several magnetic
lengths, $l=\sqrt{\hbar c/(eB)}$. The LB's in the center of the
system do approach flat Landau levels indicating that an
electron in the center does not feel the boundary. We are thus
able to study the crossing from the quantum regime in which the
electronic confinement dominates over the electron-electron  interaction to
the regime in which electrostatics plays a dominant role.
Finally we show how the formation of
a SIS can be detected in the far-infrared (FIR)
absorption spectrum of the system.

\section{Model}
We consider $N_s$ strictly two-dimensional electrons to model qualitatively
a real heterostructure where the 2DEG is confined to the lowest
electrical subband. The 2DEG is confined to a disk of radius $R$
in the 2D-plane by a potential step
\begin{equation}
      V_{\mbox{conf}}(r)=U_0
      \left[  \exp\left(\frac{R-r}{4\Delta r}\right) +1\right] ^{-1},
      \label{vconf}
\end{equation}
where $\Delta r =22\: ${\AA}. To ensure charge neutrality of the system
a positive background charge $+en_b$ resides on the disk
\begin{equation}
      n_b(r)={\bar n_s }
      \left[ \exp\left(\frac{r-R}{\Delta r}\right) +1\right] ^{-1},
      \label{nb}
\end{equation}
with the average electron density of the system given by
$\bar n_s=N_s/(\pi R^2)=\langle n_s(r)\rangle$.
In the HFA the state of each electron is
described by a single-electron Schr{\"o}dinger equation
\begin{eqnarray}
\lefteqn{\{ H^0+V_H(r)+V_{\mbox{conf}}(r)\}\psi_{\alpha}(\vec r)} \nonumber\\
     & & -\int d^2 r^{\prime}\:
      \Delta(\vec r,\vec r\: ')\psi_{\alpha}(\vec r\: ')
      =\epsilon_{\alpha}\psi_{\alpha}(\vec r) \label{schr}
\end{eqnarray}
for an electron moving in a Hartree potential
\begin{equation}
      V_H(r)={e^2\over\kappa}\int d^2 r^{\prime}
      {n_s(r^{\prime})-n_b(r^{\prime})\over |\vec r-\vec r\: '|}
      \label{vh}
\end{equation}
caused by the charge density $-e\{ n_s(r)-n_b(r)\}$,
and a nonlocal exchange potential with
\begin{equation}
      \Delta(\vec r,\vec r\: ')={e^2\over\kappa}\sum_{\beta}
      f(\epsilon_{\beta}-\mu ){\psi_{\beta}^*(\vec r\: ')
      \psi_{\beta}(\vec r)\over |\vec r-\vec r\: '|}.\label{del}
\end{equation}
The equilibrium occupation of the electronic states
is according to the Fermi distribution
$f(\epsilon_{\beta}-\mu )$ at finite temperature $T$.
The density of the electrons $n_s(r)$ is constructed from
the energy spectrum $\{ \epsilon_{\alpha}\}$ and the wave
functions $\{ \psi_{\alpha}\}$
\begin{equation}
      n_s(r)=\sum_{\alpha}|\psi_{\alpha}(\vec r)|^2
      f(\epsilon_{\alpha}-\mu), \label{ns}
\end{equation}
together with the chemical potential $\mu$.
The label $\alpha$ represents the
radial quantum number $n_r$, the angular quantum number $M$, and the
spin quantum number $s=\pm\frac{1}{2}$. $H^0$ is the single particle
Hamiltonian for one electron with spin in a constant perpendicular
external magnetic field.\cite{Landau58,Gudmundsson90:63}
A Landau band index $n$ can be constructed from the quantum numbers
$n_r$ and $M$ as $n=(|M|-M)/2+n_r$.
The Landau levels of $H^0$ with energy $E_{n,M,s}
=\hbar\omega_c(n+\frac{1}{2})+sg^*(\mu_B/\hbar )B$
are degenerate with respect to $M$
with the degeneracy $n_0=(2\pi l^2)^{-1}$ per spin orientation.
$\mu_B$ is the Bohr magneton $(e\hbar /2mc)$.
The cyclotron frequency is given by $\omega_c=eB/(mc)$.
The Hartree-Fock energy spectrum $\{ \epsilon_{\alpha}\}$ and the
corresponding wave functions $\{ \psi_{\alpha}\}$ are now found by solving
(\ref{schr})-(\ref{ns}) iteratively
in the basis
of $H^0$.\cite{Gudmundsson90:63,Gudmundsson89:517,Pfannkuche93:2244}
The chemical potential $\mu$ is recalculated in each iteration
in order to preserve the total number of electrons $N_s$.
The number of basis functions  used in the diagonalization  is chosen
such that a further increase of the subset  results in an unchanged
density $n_s(r)$.
The calculations have been repeated for several initial
conditions with different spin configuration
in order to search for the ground state of the
Hartree-Fock equations (\ref{schr})-(\ref{ns}).
The total energy of the system $E_{\mbox{tot}}$ can be found
by summing up the single electron contributions and carefully
counting the interaction energy of each electron pair
only once.\cite{Pfannkuche93:2244}

The FIR-absorption of the system is calculated as a self-consistent
linear response to an external potential\cite{Gudmundsson94:xx}
\begin{equation}
      \phi^{\mbox{ext}}(\vec r,t)={\cal E}^{\mbox{ext}}r\exp\left\{
-iN_p\varphi
      -i(\omega +i\eta )t\right\} ,
\label{phi_ext}
\end{equation}
where $\eta\rightarrow 0^+$.
$N_p=\pm 1$ corresponds to left or right circular polarization.
The small size of the system compared to the wavelength of the
external radiation makes possible to use
a electrostatic potential representing a time dependent
but spatially constant external electrical field
$\vec E^{\mbox{ext}}=-\vec\nabla\phi^{\mbox{ext}}$.
In this so-called time-dependent HFA the change
of the density matrix due to an adiabatically switched-on
total electrostatic potential
$\phi^{\mbox{sc}}$ is calculated within a linear
approximation.
The total potential consists of the external
potential and the induced potential $\phi^{\mbox{ind}}=\phi^H+\phi^F$
due to the direct and the exchange
interaction of the electrons. The induced potential in turn depends on
the density matrix, thus closing the circle and allowing for a
self-consistent evaluation of the total potential together with an
expression for the frequency dependent dielectric tensor
$\varepsilon_{\alpha\beta ,\delta\gamma}(\omega)$. The power
absorption is then calculated from the Joule heating of the system
due to $\phi^{\mbox{ext}}$
\begin{equation}
      P(\omega )=e{\cal
      E}^{\mbox{ext}}\sum_{\alpha\beta}\frac{(E_{\beta}-E_{\alpha})}
      {\hbar}\langle\beta |r|\alpha\rangle
      2\pi\delta_{M_{\beta},M_{\alpha}
      \pm N_p}\Im \left\{ f^{\alpha\beta}(\omega )
      \langle\alpha|(-e\phi^{\mbox{sc}})|\beta\rangle\right\} \, ,
\label{Pomega}
\end{equation}
where ${\cal E}^{\mbox{ext}}$ is the strength of the external field
and
\begin{equation}
      f^{\alpha \beta}(\omega ) = \frac{1}{\hbar} \, \left\{
                  \frac{f_{\beta}-f_{\alpha}}
                  {\omega + (\omega_{\beta}-\omega_{\alpha}) + i \eta}
                  \right\} \,
\end{equation}
with the Fermi distribution $f_{\alpha}=f(\epsilon_{\alpha}-\mu)$.

\section{Results}
The calculations for the box-like confinement (\ref{vconf})
are carried out with
GaAs parameters: $m^*=0.067m_e$, $\kappa =12.4$, and $g^*=-0.44$.
The occupation of the LB's is varied by changing
the number of electrons $N_s$ at a
constant strength of the magnetic field $B=3.0\: $T (we could equally
have changed the magnetic field keeping constant the number of particles).
Since the radius of the system $R\geq 1000\: ${\AA} is much larger than the
magnetic length $l\approx 148\: ${\AA} and the effective Bohr
radius $a^*_0\approx 97.9\: ${\AA} we can use as a convenient label
an effective filling factor $\nu $ describing the
occupation of the LB's in the
interior of the system. The cyclotron energy $\hbar\omega_c\approx 5.2\: $meV,
so a sufficient height of the confining potential is $U_0=60\: $meV in order
to include several LB's in the calculation.
For $B=3.0\: $T the bare spin splitting of the LB's
$E_{\mbox{Zeeman}}=(g^*\mu_B/\hbar )B\approx 0.076\: $meV
is much smaller than their separation
$\hbar\omega_c$ and corresponds to the thermal energy $k_BT$
at $T\approx 0.9\: $K.

Figure \ref{fig1} shows the HFA quasiparticle energy spectrum for four
values of  $N_s$ such that the filling factor $\nu$ (defined as the
number of occupied bands in the central
region of the box) ranges from 4 to 2.
Figure  \ref{fig1}a corresponds to the case $\nu=4$. Electrons in the
first LB ($n=0$) form a large paramagnetic
compact droplet while electrons in the second LB ($n=1$) form a smaller one.
Figures  \ref{fig1}b and  \ref{fig1}c
show clearly a large spin splitting of the LB's
due to the enhancement of $g^*$  when the 2nd LB is half filled ($\nu=3$)
and the electrons in it form a ferromagnetic compact droplet. Finally,
Fig.\ \ref{fig1}d shows the case corresponding to the droplet at $\nu=2$ when
no electrons are left in the 2nd LB.  In our case
the 2nd LB behaves in all respects like an independent, smaller quantum dot,
and its properties are identical to those studied previously.
\cite{Palacios94:5760}
However, the first LB present a more complicated behavior:
The energy spectra for six values of $N_s$ such that the chemical
potential $\mu$ lies in the neighborhood of the first LB is seen
in Fig.\ \ref{E_U}. Now the filling factor lies within the range
$1\leq \nu \leq 2$. For $N_s=42$ (Fig.\ \ref{E_U}a) both spin
bands of the first LB in the bulk region are still filled
($\mu$ is still lying between the first and second LB's but closer to the
former one). The small Zeeman energy makes
the LB's look degenerate with respect to the spin degree of
freedom. As the number of electrons is reduced to 38 (Fig.\ \ref{E_U}b)
the spin bands split up near the edge and the number of spin-down
electrons becomes smaller than that of spin-up electrons. In addition to
this splitting (which was also present in the 2nd LB in Fig.\ \ref{fig1})
we can observe two instability points (one for each
spin band) rising near the center of the system. By instability points
we mean bumps in the spin bands approaching the chemical
potential. Thus, one should expect some transition as the number of electrons
keeps changing (or the magnetic field)
and these bumps touch the  chemical potential. Before
that can happen one can even see signatures of such transitions in the density
plotted in  Fig.\ \ref{d3}. The finite temperature "reveals" a budding
spin-inversion state due to the difference in distance to
the chemical potential
between the spin-up and spin-down bands for a given position ($M$).
The onset of such a SIS takes place
when the number of particles is reduced further and those bumps cross the
chemical potential (Figs.\ \ref{E_U}c to \ref{E_U}e). Finally, the
compact droplet at $\nu=1$ is formed (Fig.\ \ref{E_U}f).

In order to understand better the formation of the SIS let us
consider a simpler but equivalent situation.
The number of LB's cannot be reduced considerably in the calculations
with the box-like confinement (\ref{vconf}) since we are using a
basis constructed of the eigenfunctions of noninteracting
electrons in an infinite system. However, by considering parabolically
confined interacting electrons and using the one-electron basis
set of such a system we are left only with band mixing due to the
electron-electron interaction.
We have thus calculated the energy spectra of a 2DEG in a parabolic
quantum dot with confinement frequency $\hbar\omega_0$
for an increasing number of LB's (from one to three) at $T=1.0\: $K.
First, we analyze the case of one LB at zero temperature.
Figure \ref{jjp} shows the
evolution with $B$ of the band structure for $N_s=30$. The spin splitting
opens up from the edge to the center of the
LB in the parabolic confinement
as we go from $\nu=2$ to $\nu=1$. Surprisingly, one can see how both spin bands
near the center of the system bend upward, and, eventually, one of them
crosses the chemical
potential.  This cannot happen for a smaller number of electrons since in that
case the $s=-\frac{1}{2}$ electrons
leave their band before the unstable point (in
the center) crosses the chemical potential.
Such central instability requires a certain size of the electronic droplet and
constitutes the initial stage of the SIS.
If we include higher LB's the spectra becomes more complicated and
the instability points of each spin band shift from each other
due to the mixing with higher LB's.
This result is presented in
Fig.\ \ref{E_para}. The total energy of the system $E_{\mbox{tot}}$
is $1789.07\: $meV for one LB, $1728.3\: $meV for two LB's, and
$1718.5\: $meV for three.
Obviously the calculation with one LB does
not represent well the ground state for the given values of
$\hbar\omega_c$ and $\hbar\omega_0$, but helps us to get an insight on the spin
instability.
{\em No twisting of the spin bands is ever seen for the calculation
with one LB.} This can be verified by checking the analytical
expressions for the matrix elements of the exchange interactions.
The additional degree of freedom introduced to the system by allowing
coupling of states of higher LB's into the lowest LB for interacting
electrons is essential in order to obtain the full richness of the
spin band structure.

The formation of the SIS invokes clear signs in the
FIR spectrum $P(\omega )$ of the 2DEG detailed in Fig.\ \ref{FIR-absorption}.
The first two subfigures show the spectrum in the Hartree approximation (HA)
and the HFA, respectively.
In the HA the exchange interaction is neglected both in the ground
state and the excited states.
A common feature is the occurrence of two strong absorption
lines, the lower one in energy corresponding to $N_p=+1$ and the higher
one corresponding to $N_p=-1$. These two lines can either be identified
as the ones corresponding to the center of mass motion predicted by the
generalized Kohn theorem for quantum dots with parabolic
confinement,\cite{Kohn61:1242,Maksym90:108,Brey89:10647,Gudmundsson91:12098,Pfannkuche94:1221}
or more appropriately here as the low energy excitation of an edge plasmon and
the 2D bulk plasmon at energy slightly higher than the cyclotron resonance
$E_c=\hbar\omega_c$.\cite{Gudmundsson94:xx}
Both approximation then show small absorption peaks above the
bulk magnetoplasmon that have been identified as absorption
due to single electron transitions.\cite{Malshukov94:2,Gudmundsson94:xx}
The spin splitting itself does not have large effects on the absorption
due to the bulk magnetoplasmon but the finer details of the corresponding
absorption peak in a parabolic quantum well have been studied by
Hembree et.\ al.,\cite{Hembree94:15197} here we shall concentrate on the
effects of the SIS. By comparing the spectra for the
two approximations at energy below the energy of the edge plasmon
we find small peaks for $N_p=-1$ that
are enlargened in the last subfigure of Fig.\ \ref{FIR-absorption}.
No such peaks are found in the HA. They are only present when the
SIS occurs and the ones with the lowest energy are
caused by single electron
transitions in the lowest LB, intra-Landau-band transitions with
$M\rightarrow M-1$ that are only possible because of the twisting
of the LB's. Corresponding absorption peaks of the opposite
polarization $N_p=+1$ can also be found in the center subfigure
at similar energy, but the peaks with $N_p=-1$ are much more
characteristic of the SIS since otherwise peaks of that
polarization are never found for low energy.
As soon as the spin Landau bands of the lowest LB cross twice a second
absorption peak appears with energy above the edge magnetoplasmon but
below the bulk plasmon. The occurrence of this second row of peaks
has to be correlated with the fact that the twisting of the lowest
and the next LB, that did mirror one another for lower $N_s$, are now
out of phase for the higher values of $N_s$ corresponding to
$\nu$ just below 2.

\section{Discussion and summary}
In a system of a confined 2DEG we have been able to
demonstrate both bulk effects and phenomena caused by the finite
size of the system, in the absence of any impurity scattering of the
electrons.
The 2D system is large enough so that the LB's approach flat
Landau levels for low values of the angular quantum number $M$.
This can be interpreted as the formation of 2D bulk states inside
the system. The ensuing singular density of states together with
the exchange interaction causes the well known oscillations of the
energy separation of the LB's with the same Landau level index $n$
but opposite spin orientations as a function of the filling factor
$\nu$. Here we have seen that the enhancement of $g^*$ occurs not
only in the LB where $\mu$ is located but in all the LB's included
in the model. Similar behavior has been established in optical
measurements of a 2DEG by Kukushkin.\cite{Kukushkin:511}

We have observed the spontaneous formation of concentric
circular regions of different spin phases when the spin
splitting of the first LB's is opening up with a decreasing $\nu$
at a low temperature. The shape of this SIS depends on the
size, shape of the system, and filling factor $\nu$,
such that the wavelength decreases as $\nu$ approaches an even
integer. The coupling of the states of higher Landau bands
into the lowest band by the Coulomb interaction of the 2DEG
is {\em essential} for the fine structure of the SIS.

Even though we have been using a restricted HFA here (total angular momentum
and spin are good quantum numbers)
different results can be attained by choosing different initial spin
configurations. In Fig.\ \ref{E_Ns30_Varg} we show three stable
states with higher energy than the ground state seen in Fig.\ \ref{E_U}c.
It is interesting to note that the state with no crossing of
spin bands is not the ground state.

The exact shape of the SIS
does strongly depend on the confining potential and, thus, also the size
of the system. As was noted earlier the LB's do not twist when $\mu$
is crossing higher LB's and the spin splitting is opening up, but the
uneven opening up produces strong modulation of the spin densities.
To exclude the possibility that numerical deficiencies are
causing the twisting of the Landau bands we have
tested the stability of the spin-density structures by increasing the
number of basis states included in the numerical calculation
and tested different schemes in attaining the convergence
of the self-consistent problem.
No visible changes in the ground state properties were observed.
On the other hand, the exact shape and formation
of the SIS does depend on the
size of the system emphasizing that we are observing a confined
spin-density wave (SDW) here.\cite{Gudmundsson94:xx,Gudmundsson94:ps}

Two possible problems associated with the HFA come to mind.
First, the HFA may lead to a ground state that is quite different
from the physical one due to the strong exchange force that may be
reduced in better approximations where higher order correlation effects
or impurity broadening to a high order are included.\cite{Uenoyama89:11044}
It is thus, very reassuring that this type of spin inversion and
formation of a SDW has been observed in models employing
the local density approximation (LDA) where the SDW has
been observed for different approximations of the correlation
effects.\cite{Hembree93:9162} The on-set of the SDW is
also found to depend on the amount of collision broadening of the LB's,
but neither the broadening nor the correlation effects
prevent it.\cite{Hembree93:9162} The spatial correlation of the
2DEG in two approaching finite-size layers for the common filling factor
of unity is quite similar to the formation of the SIS here.
The layer index can be treated as isospin for vanishing separation
and the numerical diagonalization of the
many-electron Hamiltonian in a large subspace of noninteracting
many-electron states includes, in principle, all correlation effects
in the model to a high degree of accuracy.\cite{Palacios95:1769}

An important difference of the present SIS
in the two-dimensional plane to the SDW parallel to $\vec B$ investigated
by Brey and Halperin \cite{Brey89:11634} is the fact that the
wavelength of the present modulation varies strongly with
$\nu$. This is caused by the strong dependence of the effective
interaction, or the screening, in the 2D plane on
$\nu$.\cite{Gudmundsson90:63,Wulf88:4218,Labbe88:1373,McEuen92:11419}
The SDW  found by Brey and Halperin has strong
reassemblance with the more ``traditional one'' known
in 1D electroninc systems.\cite{Gruener:xx} The notation SIS
is, therefore, used here to emphazise this difference.

The region of filling factors when the electrons
are not fully spin polarized yet ($1\leq\nu\leq 2$)
but the system has not entered the
regime of the integer quantum Hall effect with the lowest LB
filled ($\nu=1$) has attracted much interest lately. It has been shown that
in absence of Zeeman energy the lowest
energy charged excitations at $\nu =1$ are skyrmions, spin textures
with a unit winding number in two
dimensions.\cite{Sondhi93:16419,Fertig94:11018}
At large $g$ the quasi-particles, analogous to the single particles, have
unit charge $\pm e$ and spin half, $s=\pm 1/2$, but as $g$ is reduced
to zero the excitation gap survives and the size of the quasi-particles
diverges with the spin becoming macroscopic - skyrmions.\cite{Fertig94:11018}
This effect has also been studied in double-layered electron systems
when the distance between the layers, each having no spin degree of freedom,
is reduced since these models can be mapped directly onto the spin system
identifying the layer index as an
isospin.\cite{Palacios95:1769,Moon95:5138}
It has also been found that
these spin textures might eventually dominate the ground state properties
at filling factors
$1\leq\nu\leq 2$.\cite{Palacios94:5760,Palacios95:1769,Schmidt95:5570}
The SIS's found in the present work are
{\em not} related to the skyrmions observed in such regime of filling factors,
but the skyrmions and the  SIS's
may coexist, which emphasizes the
very complex and interesting structure of the 2DEG in such a regime.

The spin-density modulation was found to cause clear signs
in the FIR-absoption of the confined 2DEG. The signs may be
weak since they partly reflect single-electron transitions rather than
collective oscillations and they may be in the low frequency part
of the spectrum most difficult to measure, but the final
word about the appropriatness of the HFA or the LDA for the
current model will come from experiments.

\acknowledgements
The authors are indebted to R.\ R.\ Gerhardts, D.\ Pfannkuche,
A.\ H.\ MacDonald, M.\ Ferconi, G.\ Vignale,
and G.\ P{\'a}lsson for fruitful discussion.
This research was supported in part by the Icelandic Natural Science
Foundation, the University of Iceland Research Fund,
NATO collaborative research Grant No. CRG 921204, NFS contract
NSF-DMR9416906, and, for
one of the authors (J.J.P.), by a NATO postdoctoral fellowship.
\bibliographystyle{prsty}

%\newpage
%
%
\begin{figure}
\caption{The ground state HFA energy spectra and chemical potential $\mu$
         (horizontal line) for (a) $N_s=82$, (b) $N_s=62$,
         (c) $N_s=52$, and (d) $N_s=48$ electrons in the system.
         $T=10.0\: $K, $R=1000\: ${\AA},
         $U_0=60\: $meV, and $B=3.0\: $T.
         Crosses represent the $s=+\frac{1}{2}$ electrons and
         diamonds represent $s=-\frac{1}{2}$.
         GaAs bulk parameters: $m^* = 0.067m_0$, $\kappa = 12.4$,
         $g^* = -0.44$.
}
\label{fig1}
\end{figure}

\begin{figure}
\caption{The ground state energy spectra and chemical potential $\mu$
         (horizontal line) for (a) $N_s=42$, (b) $N_s=38$,
         (c) $N_s=34$, (d) $N_s=30$, (e) $N_s=26$, and (f)
         $N_s=22$ electrons in the system.
         $T=4.0\: $K, $R=1000\: ${\AA},
         $U_0=60\: $meV, and $B=3.0\: $T.
         Crosses represent the $s=+\frac{1}{2}$ electrons and
         diamonds represent $s=-\frac{1}{2}$.
}
\label{E_U}
\end{figure}
\begin{figure}
\caption{The ground state electron density $n_s(r)$ for $s=-1/2$ (solid) and
         $s=+1/2$ (dashed) in the case of $T=4.0\: $K.
         Other parameters are as in Fig.~\protect\ref{fig1}.
}
\label{d3}
\end{figure}
\begin{figure}
\caption{HFA ground state energy spectra and chemical potential
         $\mu$ for $30$ electrons in a parabolic confinement
         potential at $T=0\: $K for (a) $\hbar\omega_c=3.0$ meV,
         (b)$\hbar\omega_c=5.0$ meV, and (c) $\hbar\omega_c=7.0$ meV. Only
         one LB is considered. Confinement
         frequency $\hbar\omega_0=5.0\: $meV.
         Crosses represent the $s=+\frac{1}{2}$ electrons and
         diamonds represent $s=-\frac{1}{2}$.
         Other parameters are as in Fig.~\protect\ref{fig1}.
}
\label{jjp}
\end{figure}
\begin{figure}
\caption{HFA ground state energy spectra and chemical potential
         $\mu$ for $30$ electrons in a parabolic confinement
         potential at $T=1.0\: $K.
         $1$ LB is used for the calculation of the top
         subfigure, $2$ for the center one, and $3$
         for the bottem one.
         $\hbar\omega_c=8.0\: $meV, and the confinement
         frequency $\hbar\omega_0=4.0\: $meV.
         Crosses represent the $s=+\frac{1}{2}$ electrons and
         diamonds represent $s=-\frac{1}{2}$.
         Other parameters are as in Fig.~\protect\ref{fig1}.
}
\label{E_para}
\end{figure}
\begin{figure}
\caption{The FIR-absorption $P(E)$ vs.\ $E/E_c$ ($E_c=\hbar\omega_c$)
         and the number of electrons $N_s$
         for $N_p=\pm 1$ for HA (left), HFA (center), and for $N_p=-1$
         in the HFA (right).$T=4.0$ K and
         other parameters are as in Fig.~\protect\ref{fig1}.
}
\label{FIR-absorption}
\end{figure}
\begin{figure}
\caption{Energy spectra and chemical potential $\mu$ (horizontal line)
         of several stable excited states for $N_s=34$.
         Same parameters as in
         Fig.~\protect\ref{E_U}.
}
\label{E_Ns30_Varg}
\end{figure}
\end{document}